\documentclass[twocolumn,showpacs,preprintnumbers,amsmath,amssymb,prb,aps]{revtex4}
\usepackage{graphicx}
\usepackage{bm}

\begin{document}

\title{Bose-glass to Superfluid transition in the three-dimensional Bose-Hubbard Model}

\author{Peter Hitchcock}
	\email{hitchpa@mcmaster.ca}
\author{Erik S. S\o rensen}
	\email{sorensen@mcmaster.ca}
\affiliation{Department of Physics and Astronomy, McMaster University, Hamilton, ON, L8S 4M1 Canada}

\date{\today}

\begin{abstract}
We present a Monte Carlo study of the Bose-glass to superfluid transition in the three-dimensional
Bose-Hubbard model. Simulations are performed on the classical (3~+~1)~dimensional link-current
representation using the geometrical worm algorithm. Finite-size scaling analysis (on lattices as
large as 16x16x16x512 sites) of the superfluid stiffness and the compressibility is consistent with
a value of the dynamical critical exponent $z=3$, in agreement with existing scaling and
renormalization group arguments that $z=d$. We find also a value of $\nu=0.70(12)$ for the
correlation length exponent, satisfying the relation $\nu \ge 2/d$.  However, a
detailed study of the correlation functions, $C(r,\tau)$, at the quantum critical point are not
consistent with this value of $z$. We speculate that this discrepancy could be due to the fact that
the correlation functions have not reached their true asymptotic behavior because of the relatively
small spatial extent of the lattices used in the present study.
\end{abstract}


\maketitle

\section{\label{sec:intro}Introduction}
Quantum phase transitions as they occur in models comprised of bosons have been the focus of
considerable interest lately. Most notably, experiments by Greiner~\textit{et
al.}~\cite{greiner:2002a:nature} have observed the transition between a Mott insulating (MI) phase
and a superfluid phase in an optical lattice loaded with $^{87}$Rb atoms.  The observed phase
transition between the superfluid and the insulating phase is thought to share the universal
properties of a variety of physical systems, including He$^4$ in porous
substrates~\cite{crowell:1997:prb}, Josephson-junction arrays~\cite{Zant} as well as thin
superconducting films~\cite{Liu,Markovic,goldman}.  While disorder can be neglected in the
experiments by Greiner \textit{et al.}~\cite{greiner:2002a:nature}, it clearly plays a central role
in several of the experiments just mentioned. As outlined in the seminal paper by Fisher \textit{et
al.}~\cite{fisher:1989:prb} the statics as well as the dynamics of the quantum phase transitions
occurring in the Bose-Hubbard model are strongly influenced by disorder and a new insulating glassy
phase, the Bose-glass phase, should occur. The transition of interest in this case is directly
between the superfluid (SF) and the Bose-glass (BG). Recent experiments showing that this disordered
transition can also be studied using optical lattices~\cite{Lye} has therefore created considerable
excitement.  Previous theoretical studies~\cite{Giamarchi,Trivedi,Sorensen,Batrouni,Makivic,Jaksch}
have mainly focused on the transition as it occurs in one or two dimensions and very recent
theoretical works~\cite{Roth,Damski}, addressing directly the experimental situation relevant for
optical lattices, have also mainly analyzed this case.  In light of the experiments by
Lye~\textit{et al.}~\cite{Lye,Fort}, we consider in the present study the transition as it occurs in
three dimensions in the presence of short-range interactions. In particular we determine the
dynamical critical exponent, $z$, characterizing this transition in $d=3$.

One hall-mark of the Mott insulating phase is that it is incompressible. In contrast to this, both
the superfluid and the Bose-glass phase are compressible and have a finite non-zero compressibility,
$\kappa\neq 0$.  Close to  the quantum critical point between the Bose-glass and the superfluid
scaling arguments~\cite{fisher:1989:prb}, based on generalized Josephson relations, then show that:
\begin{equation}
\kappa\sim\delta^{\nu(d-z)},
\end{equation}
where $\delta$ describes the distance to the critical point in terms of the control parameter.
Since both the BG and SF phases are compressible it then seems natural to expect the compressibility
to be finite {\it also} at the critical point and one must then conclude that
\begin{equation}
z=d.
\end{equation}
It can in fact be shown that assuming the compressibility either diverges or tends to zero at the
BG-SF critical point leads to implausible behavior.  In Ref.~\onlinecite{fisher:1989:prb} the
relation $z=d$ was therefore argued to hold not only below the upper critical dimension but for {\it
any} dimension $d\geq 1$.  This result implies that there is not an upper critical dimension for
this transition in any conventional sense.  In $d=1$ analytical work~\cite{Giamarchi} strongly
supports the conclusion that $z=d=1$ and numerical work~\cite{Sorensen,Batrouni,alet:2003a:pre} in
$d=2$ also find $z=d=2$.  Long-range interactions are likely to yield a different
$z$~\cite{fisher:1989:prb}, but we shall not be concerned with that case here.  Based on
Dorogovtsev's~\cite{Dorogovtsev} double-$\epsilon$ expansion, an alternative scenario has been
proposed~\cite{weichman:1989:prb,mukhopadhyay:1996:prl} in which the critical exponents jump
discontinuously to their mean field values at the critical dimension $d_c=4$.  In this scenario, for
$d>d_c$ two stable fixed points are present: the gaussian fixed point is stable at {\it weak}
disorder with the random fixed point remaining stable at {\it strong} disorder. Below the critical
dimension, $d<d_c$, only the random fixed point is stable. It should be noted that in order to
obtain sensible answers from the double-$\epsilon$ expansion an ultraviolet frequency cutoff
$\omega_\Lambda$ has to be introduced~\cite{weichman:1989:prb}---a procedure which seems difficult
to justify. This was later remedied upon~\cite{mukhopadhyay:1996:prl} still within the framework of
the double-$\epsilon$ expansion.  It is also possible to question the validity of any
$\epsilon$-expansion approach on the grounds that the existence of an upper critical dimension for
the disordered transition is implicitly assumed.  One might then ask if bounds on the upper critical
dimensions exist. Indeed, using the exact inequality~\cite{chayes:1986:prl} 
\begin{equation}
\nu \geq 2/d, 
\end{equation}
valid for stable fixed points in the presence of correlated disorder, one immediately sees that the
requirement that $\nu=1/2$ at the upper critical dimension, $d_c$, yields $d_c\geq 4$.  Since the
existence of an upper critical dimension is debatable it is then natural to instead focus on the
lower-critical dimension, which is well established.  Indeed, it is also possible to develop an
expansion away from the lower-critical dimension ($d_l$ = 1)~\cite{herbut:1998:prb,
herbut:2000:prb}.  Using this approach the dynamical critical exponent is exactly $z=d$ and the
correlation length exponent $\nu$ can be calculated to second order in $\sqrt{d - 1}$ yielding good
agreement with experimental and numerical results in $d=2$.  The onset of mean field behavior, if
any, is therefore at best unconventional in this model and still a matter of debate.  In particular,
it would be valuable to know if the relation $z=d$ continues to hold in dimensions higher than
$d=2$.

In the present work, we present a Monte Carlo study of the three-dimensional Bose-glass to
superfluid transition at strong disorder. Our focus is reliable estimates of the dynamical critical
exponent $z$ in order to test the relation $z=d=3$.  As outlined above, much of the numerical work
to date has focused on the one- and two-dimensional cases, which are less demanding computationally.
However, recently a very efficient geometric worm algorithm~\cite{alet:2003a:pre,alet:2003b:pre} has
been developed for the study of bosonic phase transitions making significantly larger system sizes
available.  Even though the geometrical worm algorithm we use has proven to be very efficient for
dealing with large lattices in two dimensions (also in the presence of disorder), we were only able
to study a relatively limited number of system sizes in $d=3$ that were large enough to avoid
finite-size effects but small enough to properly equilibrate with a feasible amount of computational
effort.  In the present study therefore, we focus exclusively on the three dimensional case. We
leave for future study the case of $d=4$ that would be of considerable interest for the approach to
mean field suggested by Weichman and collaborators~\cite{weichman:1989:prb,mukhopadhyay:1996:prl}.

It has been suggested that in the vicinity of commensurate values of the chemical potential disorder
is weakly irrelevant~\cite{zhang:1992:prb,singh:1992:prb,kisker:1997:prb,pazmandi:1998:prb}.  In
this case a direct MI-SF transition could occur even in the presence of weak disorder.  However,
recent high precision numerical work~\cite{prokof'ev:2004:prl} reached the opposite conclusion for
$d=2$, although very large system sizes were needed in order to show this.  Since we are mainly
interested in the BG-SF transition, we minimize any cross-over effects by performing all of our
calculations at a value of the chemical potential where in the absence of any disorder the system is
in the superfluid phase for any non-zero hopping and there is {\it no} MI-SF
transition~\cite{fisher:1989:prb}. The BG-SF transition we observe is therefore induced by the
disorder and corresponds in the RG sense to a random fixed point.

The paper is outlined as follows: the Bose-Hubbard model is introduced below in further detail,
including the mapping to a $(d + 1)$ dimensional classical model on which we perform our study.
Section \ref{sec:scaling} outlines the scaling theories necessary to extract the critical exponents.
Section \ref{sec:numeric} details the numerical techniques and discusses the difficulties we
encountered in properly equilibrating our simulations. Finally, we present our results in section
\ref{sec:results}.

\section{\label{sec:model}Model} 
We begin with the Bose-Hubbard Hamiltonian, including on-site disorder in the chemical
potential:~\cite{fisher:1989:prb}
\begin{equation} 
H_{\mathrm{BH}} = \sum_{{\bf r}}\Big(\frac{U}{2}\hat{n}_{{\bf r}}(\hat{n}_{{\bf r}} - 1) - \mu_{{\bf r}}\hat{n}_{{\bf r}}\Big) - 
\frac{t}{2}\sum_{\langle {\bf r}, {\bf r}'\rangle}({\hat{\bf \Phi}^\dagger_{\bf r}\hat{\bf \Phi}_{\bf r'}} + \mathrm{H.c.}). 
\end{equation}
Here $ {\hat{\bf \Phi}_{\bf r}^\dagger} $ and $ {\hat{\bf \Phi}_{\bf r}} $ are boson creation and
annihilation operators at site $ {\bf r} $, and $ \hat{n}_{{\bf r}} = {\hat{\bf \Phi}_{{\bf
r}}^\dagger}{\hat{\bf \Phi}_{\bf r}} $ is the number operator. The on-site repulsive interaction $U$
localizes the bosons and competes with the delocalizing effects of the tunneling coefficient $t$.
The random chemical potential $\mu_{{\bf r}}$ is distributed uniformly on $ [\mu - \Delta, \mu +
\Delta] $.  As noted by Damski~\textit{et al.}~\cite{Damski}, the introduction of a random potential
in an optical lattice will also generate randomness in the hopping (tunneling) term, $t$. However,
this type of disorder can be ignored~\cite{Damski}.  The phase diagram~\cite{fisher:1989:prb} for
the pure model consists of a superfluid phase at high $t/U$ which is unstable to a series of
Mott-insulating regimes centered at commensurate densities at low $t/U$. In the presence of
disorder, the Bose-glass phase stabilizes between the insulating and superfluid phases. 

Following standard methods~\cite{Sorensen,wallin:1994:prb}, by integrating out amplitude
fluctuations of the Bose field to second order, we transform the Bose-Hubbard Hamiltonian into an
effective classical Hamiltonian in ($d$~+~1)~dimensions well suited for Monte Carlo study:
\begin{equation} 
\label{eq:Hlc} 
H = \frac{1}{K}\sum_{({\bf r}, \tau)}\Big[\frac{1}{2}{\bf J}^2_{({\bf r}, \tau)} - \mu_{{\bf r}}J^\tau_{({\bf r}, \tau)}\Big].  
\end{equation}
The integer currents ${\bf J}_{({\bf r}, \tau)}$ are defined on the bonds of the lattice and obey a
divergenceless constraint $\nabla {\bf J}_{({\bf r}, \tau)} = 0$. The resulting current loops are
interpreted in this context as world lines of bosons~\cite{Sorensen,wallin:1994:prb} and represent
fluctuations from an average, non-zero density. The coupling $K$ acts as an effective classical
temperature and drives the phase transition: at low $K$ the repulsive short range interaction $U$
dominates and the system is insulating, while at high $K$ the hopping $t$ dominates and the system
is superfluid.  In this transformation amplitude fluctuations of the boson fields are integrated out
to second order. However, these fluctuations are not expected to affect the universal details of the
transition. 

The advantage of the formulation of the Bose-Hubbard model in terms of the link-currents,
Eq.~(\ref{eq:Hlc}), is that an extremely efficient worm algorithm~\cite{alet:2003a:pre} is available
for this model. This worm algorithm can also be directed~\cite{alet:2003b:pre} even in the presence
of disorder. However, the memory overhead associated with using a directed algorithm is prohibitive
for the present study due to the presence of disorder and the high spatial dimension. We have
therefore exclusively used the more standard formulation of the algorithm~\cite{alet:2003a:pre}.

\section{Method} 
\subsection{\label{sec:scaling}Scaling} 
The BG-SF transition is a continuous quantum phase transition, and is characterized by two diverging
length scales: a spatial ($\xi$) and a temporal ($\xi_\tau$) correlation length, related by the
dynamical critical exponent $z$:
\begin{equation} 
\xi_\tau \sim \xi^z \sim (\delta^{-\nu})^z, 
\end{equation} 
with $\delta = (K - K_c)/K_c$. These form the basis of the scaling theory used to analyze our data.

The two observables of primary interest are the superfluid stiffness $\rho$ and the compressibility
$\kappa$.  The superfluid stiffness $\rho$ (proportional to the superfluid density) is defined in
terms of the change in free energy associated with a twist in the spatial boundary conditions. As
for the compressibility, the critical behaviour of $\rho$ can be derived as a generalization of the
Josephson scaling relations for the classical transition~\cite{fisher:1989:prb,Sorensen,%
wallin:1994:prb}, and scales with the correlation length as
\begin{equation}
\rho \sim \xi^{-(d + z - 2)}.
\end{equation}

The compressibility can similarly be found as a response to a twist applied to the temporal boundary
conditions, and is found to scale as~\cite{fisher:1989:prb,Sorensen,wallin:1994:prb}
\begin{equation}
\kappa \sim \xi^{-(d - z)},
\end{equation}
leading to the relation $z=d$ previously mentioned.

The first step in the study of the critical properties of the Bose-Hubbard model is a precise
determination of the location of the critical point through finite-size scaling analysis. The
presence of two correlation lengths implies that finite-size scaling functions will have two
arguments: $f(L/\xi, L_\tau/\xi_\tau)$. Hence
\begin{equation}
\rho = \xi^{-(d + z - 2)}f(L/\xi, L_\tau/\xi_\tau),
\end{equation}
or, by appropriately scaling the arguments and requiring that the stiffness remain finite at the
critical point for a system of finite size,
\begin{equation}
\label{eq:pscale}
\rho = \frac{1}{L^{d + z - 2}}\bar{\rho}(L^{1/\nu}\delta, L_\tau/L^z).
\end{equation}
This complicates the scaling analysis as we must work in a two-dimensional space.  The first
approach is to work at lattice sizes whose temporal sizes scale with the exponent $z$; that is to
work at a fixed aspect ratio:
\begin{equation}
\alpha = L_\tau/L^z.
\end{equation}
The quantity:
\begin{equation}
L^{d + z - 2}
\rho(L^{1/\nu}\delta,\alpha=\frac{L_\tau}{L^z})
\label{eq:rhoscale}
\end{equation}
should then be a universal function of $\alpha$ at $K_c$. We can hold the aspect ratio constant by
working with systems of dimension $L^d\times\alpha L^z$.  If an initial estimate for the dynamical
critical exponent $z$ is available, the critical point can be located by plotting $L^{d+z-2}\rho
(L^{1/\nu}\delta,\alpha)$ versus $K$ for several different linear system sizes $L$. If the correct
value of $z$ is used, these curves will all intersect at the critical point.  This unfortunately
requires that the initial estimate for the value of $z$ be made before simulations are run.  

Alternatively, with an estimate of $K_c$, the first argument of Eq. \ref{eq:rhoscale} can be held
constant. Curves of $\rho L^{d + z - 2}$ plotted for different $L$ against the ratio $L_\tau/L^z$
should then collapse for the correct value of $z$~\cite{guo:1994:prl,GuoPRB}.  We use both
approaches, with the hope that a consistent picture emerges.

Once $K_c$ has been located by the above method we can proceed to study the behavior of the
compressibility at the critical point.  If indeed $z$ is equal to $d$, as described in the
introduction, the compressibility should be roughly constant as $K$ varies through $K_c$, and should
not show any dependence on $L$.  In particular $\kappa$ should neither diverge nor go to zero at
$K_c$.

We also consider the particle-particle correlation function $C({\bf r}, \tau)$, which is
expected~\cite{fisher:1989:prb} to decay asymptotically as
\begin{equation}
C({\bf r}, \tau = 0) \sim r^{-y_r},\ C(\bm{r} = 0, \tau) \sim \tau^{-y_\tau}
\end{equation}
with exponents given by
\begin{eqnarray}
y_r & = & d + z - 2 + \eta, \nonumber \\
y_\tau & = & (d + z - 2 + \eta)/z.
\end{eqnarray}
In addition to defining the exponent $\eta$, with reliable estimates of the correlation functions
this would in principle provide for an independent calculation of the dynamical critical exponent
$z$.  Fisher \textit{et al.}~\cite{fisher:1989:prb} also derive bounds for $\eta$:
\begin{equation}
2 - (d + z) < \eta \le 2 - d.
\label{eq:eta}
\end{equation}
The lower bound arises from the requirement that the correlation functions decay, and the upper
bound is argued from the scaling of the density of states in the Bose-glass and superfluid phases.
These can be simply stated as bounds on the exponents $y_r$ and $y_\tau$:
\begin{equation}
0 < y_\tau \le 1,\ 0 < y_r \le z.
\end{equation}

\begin{figure}[t,floatfix]
\includegraphics[clip,width=8cm]{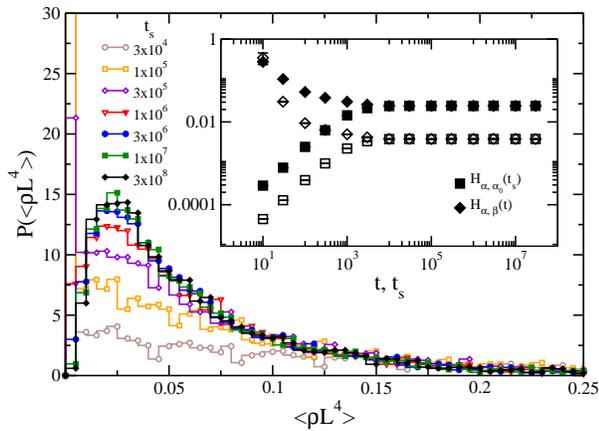}
\caption{Main panel: histograms of $\langle \rho L^4\rangle$ for a set of 1000 disorder realizations
			on a system of size 8x8x8x64 at our estimate of $K_c = 0.190(1)$.  Curves show the evolution of the
			distribution $P(\langle \rho L^4\rangle)$ as a function of the number of Monte Carlo sweeps $t_s$
			performed after equilibration on each disorder realization.  The peak in the distributions at
			$\langle \rho L^4\rangle = 0$ persists for many times the relaxation time, as measured by the
			convergence of the Hamming distances, while the broader peak near $\langle \rho L^4\rangle = 0.025$
			grows.  Inset: Hamming distances calculated on the same set of disorder realizations. $H_{\alpha,
			\beta}$ is plotted against the total number of sweeps performed, $t$. $H_{\alpha, \alpha_0}$ is
			plotted against the number of sweeps performed after equilibration, $t_s$. Open symbols are
			calculated in terms of spatial currents.  Solid symbols are calculated in terms of temporal
			currents. }
\label{fig:equilib}
\end{figure}

The final exponent we calculate is the correlation length exponent $\nu$. Taking the derivative of
(\ref{eq:pscale}) with respect to $K$ we see that
\begin{equation}
L^{d+z-2}\frac{d\rho}{dK} = L^{1/\nu}\bar{\rho}'(L^{1/\nu}\delta, L_\tau/L^z).
\end{equation}
Plotting this derivative against $L$ at $K_c$ should yield a power law with exponent $1/\nu$. The
crossing data $\rho L^{d + z - 2}$ calculated for different system sizes with a fixed aspect ratio,
$\alpha$, should also collapse to a single curve when plotted against $L^{1/\nu}\delta$.

\subsection{\label{sec:numeric}Numerical Method}
Both the superfluid stiffness and the compressibility can be calculated in terms of the link-current
winding numbers~\cite{Sorensen,wallin:1994:prb} $n_{\gamma} = L_\gamma^{-1}\sum_{{\bf r},
\tau}J^{\gamma}_{{\bf r}, \tau}$ in each direction $\gamma = x, y, z, \tau$ (here $L_{x, y, z} =
L$). The superfluid stiffness is related to a twist in the spatial boundary conditions and so can be
calculated in terms of the spatial winding numbers: 
\begin{equation}
\label{eq:stiffness}
\rho = \frac{1}{L^{d - 2}L_\tau}\big[\langle n_{\gamma=x,y,z}^2 \rangle\big]_{av}.
\end{equation}
We denote thermal averages by $\langle\mathcal{O}\rangle$ and disorder averages by
$\big[\mathcal{O}\big]_{av}$. Similarly, the compressibility is associated with a twist in the
temporal boundary conditions and is defined in terms of the temporal winding numbers:
\begin{equation}
\kappa = \frac{L_\tau}{L^d}\big[\langle n_{\tau}^2 \rangle - \langle n_{\tau}\rangle_\alpha\langle n_{\tau}\rangle_\beta\big]_{av}.
\end{equation}
Since this expression contains the disorder average of the square of a thermal average, the
systematic error is reduced by calculating this average on two independent lattices $\alpha$ and
$\beta$ with the same disorder realization~\cite{bhatt:1988:prb,hpcs2006}.

\begin{figure}[t,floatfix]
\includegraphics[clip,width=8cm]{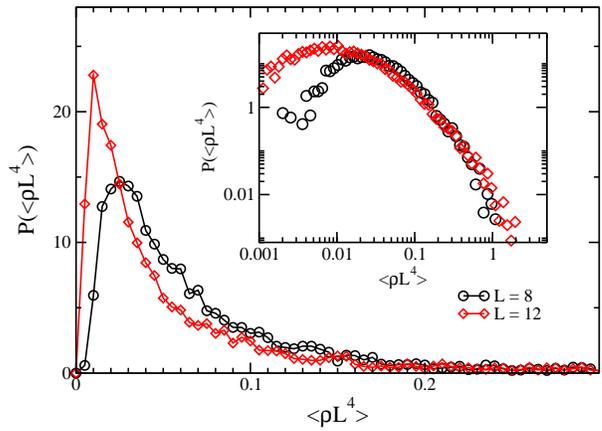}
\caption{Distributions of $\langle\rho L^4\rangle$ for $L = 8$ and 12 at $K_c$, plotted on linear
			(main panel) and logarithmic (inset) axes. The average of each distributions is close to $[\langle
			\rho L^4\rangle]_{av} = 0.08$, which is significantly higher than the typical values of the
			distributions, indicating that the broad tail of the distribution must be well sampled to obtain an
			accurate estimate of the true average. } 
\label{fig:pdist} 
\end{figure}

We are also interested in the derivative of $\rho$ with respect to the coupling. This can be
calculated thermodynamically from the stiffness and total energy $E$:
\begin{equation}
\frac{d\rho}{dK} = \frac{1}{K^2}\big[\langle\rho E\rangle - \langle\rho\rangle_\alpha\langle E\rangle_\beta\big]_{av}.
\end{equation}
This was found to produce better estimates than numerically differentiating $\rho$ in the
two-dimensional case~\cite{alet:2004:prb}. 

Finally, since the construction of each `worm' is essentially equivalent to propagating a boson
through the lattice, it is possible to calculate the particle-particle correlation function directly
from the behaviour of the algorithm. Details can be found in Ref.~\onlinecite{alet:2003b:pre}.

As always, the system must be be run for $t_0$ Monte Carlo sweeps at each disorder realization to
ensure that equilibrium has been reached before beginning to sample the generated configurations. To
confirm this, two simulations are carried out simultaneously on lattices with the same disorder
realization but different initial configurations (we set all of the currents in each direction to a
different integer constant for $\alpha$ and $\beta$). These initial configurations are far from
equilibrium.  It is useful to define `Hamming distances' between different current configurations in
order to measure the relaxation time of the algorithm (this is done in the spirit of
Ref.~\onlinecite{wallin:1994:prb}, though the definitions used here are slightly different). We
define the Hamming distance between the two lattices $\alpha$ and $\beta$ after performing a total
of $t$ Monte Carlo sweeps on their initial configurations: 
\begin{equation}
\label{eq:hamming}
H_{\alpha, \beta}^{\nu = x, \tau}(t) = \frac{1}{L^dL_\tau}\sum_{({\bf r}, \tau)}
\big[J^\nu_{\alpha,({\bf r}, \tau)}(t) - J^\nu_{\beta,({\bf r}, \tau)}(t)\big]^2.
\end{equation}
Similarly, we define the Hamming distance between the configuration of lattice $\alpha$ at the sweep
$t_0$ where we begin to sample the generated configurations (denoted $\alpha_0$) and the
configuration of $\alpha$ after a further $t_s = t - t_0$ sweeps:
\begin{displaymath}
H_{\alpha, \alpha_0}^{\nu = x, \tau}(t_s) = \frac{1}{L^dL_\tau} \sum_{({\bf r}, \tau)}
\big[J^\nu_{\alpha,({\bf r}, \tau)}(t_0 + t_s) - J^\nu_{\alpha,({\bf r}, \tau)}(t_0)\big]^2.
\end{displaymath}
In the present study, $t_0$ is chosen prior to running the simulations and is held constant. 

For the sake of simplicity, we define a Monte Carlo sweep to be the construction of a single worm;
while not ideal for comparing its characteristics to other algorithms, this is sufficient for our
discussion here. Since the initial configuration of $\alpha$ and $\beta$ are different, $H_{\alpha,
\beta}^{\gamma}(t)$ will be large at the beginning of the simulation ($t$ small). Conversely, since
shortly after $t_0$ the configuration of $\alpha$ will not have changed substantially, $H_{\alpha,
\alpha_0}^{\gamma}(t_s)$ will be small.  If $t_0$ has been chosen larger than the relaxation time of
the algorithm, then for sufficiently large values of $t$ and $t_s$ the configurations of $\alpha_0$,
$\alpha$, and $\beta$ will be independent, equilibrated states, and thus the two distances should
converge. The inset of figure \ref{fig:equilib} shows this approach to equilibrium on an 8x8x8x64
lattice with $t_0 = 3 \times 10^7$ at our estimate of $K_c$. The distances converge after $t \approx
10^5$, indicating that the relaxation time for this system is approximately thirty thousand sweeps.

\begin{figure}[t,floatfix]
\includegraphics[clip,width=8cm]{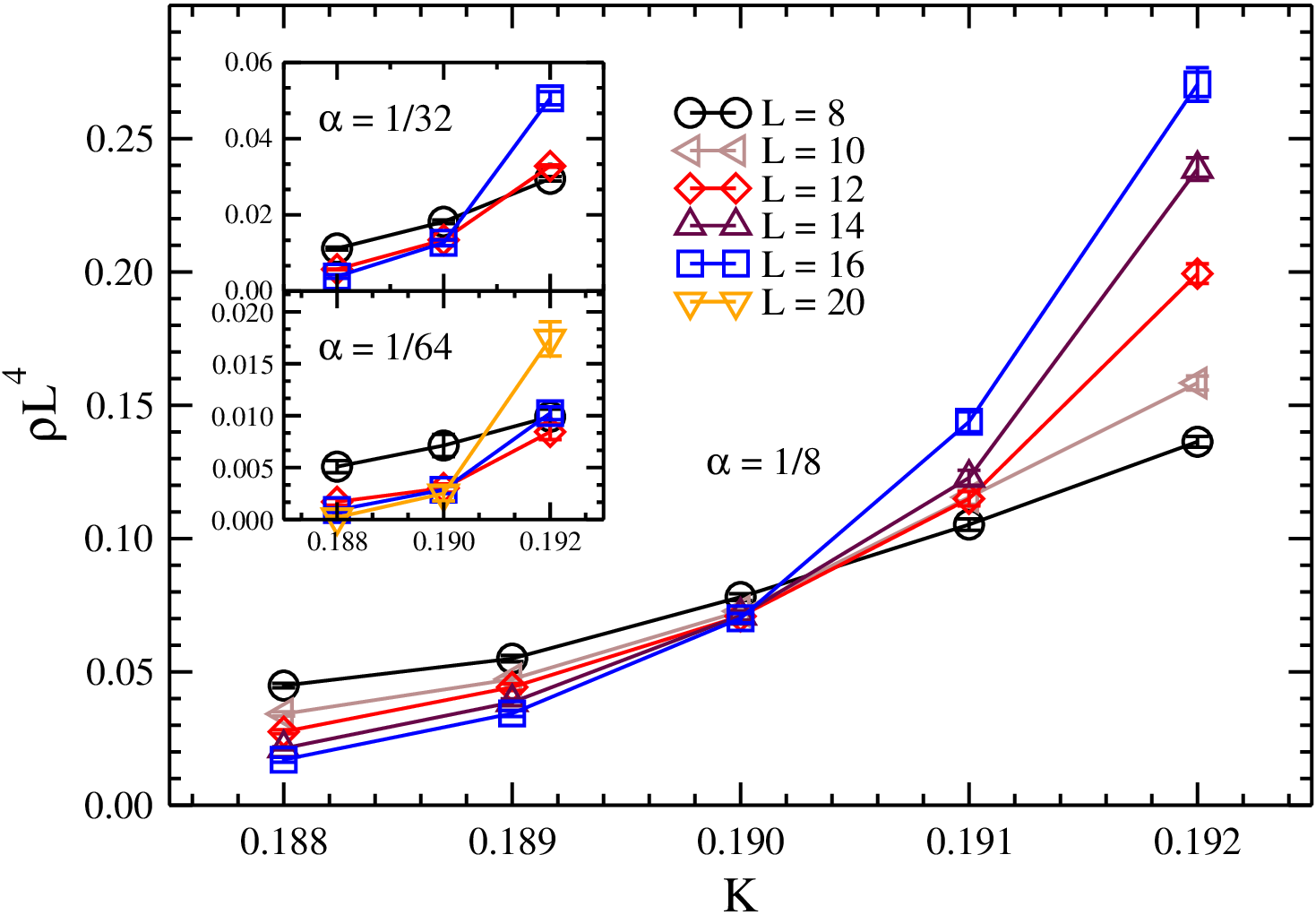}
\caption{The main panel shows $\rho L^{d + z - 2} = \rho L^4$ plotted
			versus $K$ for lattices of size $L_\tau = \frac{1}{8}L^3$ assuming $d=z=3$.  The lines are guides to
			the eye.  The crossing gives an estimate of the critical point $K_c = 0.190(1)$ and is consistent
			with $z$ = 3. The two insets show equivalent results with two different aspect ratios $\alpha=1/32$
			and $\alpha=1/64$.} 
\label{fig:cross}
\end{figure}

A further complication in the calculation of the disorder averages was encountered due to the
discrete nature of the winding numbers which are used to calculate $\rho$ and $\kappa$. Since most
disorder realizations have a small but finite superfluid stiffness at $K_c$, many independent
configurations of $n_\gamma$ must be generated for each disorder realization to achieve a reliable
estimate. Figure \ref{fig:equilib} shows the distribution of $\langle \rho L^4 \rangle$ as a
function of the number of sweeps $t_s$ performed after equilibration on each realization of the
disorder. For $t_s \approx t_r$, there are many realizations for which no configuration is generated
with a finite winding number.  Only after running for much longer at each realization are the true
features of the distribution resolved. Under-sampling these realizations runs the risk of
underestimating the disorder average.  This issue is discussed at further length in
Ref.~\onlinecite{hpcs2006}.

The computational demands of equilibration grow quickly with system size. Typical systems we
studied, of linear size $L = 10$ to $L = 14$, required up to $3\times 10^5$ sweeps to relax and up
to a further $1\times 10^8$ sweeps for the distributions of $\langle\rho L^4\rangle$ to equilibrate.
For $L=16$, nearly $1\times 10^6$ sweeps were required to relax and we were unable to run
simulations for sufficiently long to see the distributions equilibrate. Some results for $L = 16$
are shown below as they demonstrate consistent scaling. Figure \ref{fig:pdist} shows the
equilibrated distributions of $\langle \rho L^4 \rangle$ for two lattice sizes, $L = 8$ and 12. They
show the broad tail which necessitates running on at least $10^3$ disorder realizations. Moreover,
at least for the system sizes we were able to study, the distributions do not narrow as $L$
increases.  Hence, the system is likely not self-averaging~\cite{aharony:1996:prl}.

\begin{figure}[t,floatfix]
\includegraphics[clip,width=8cm]{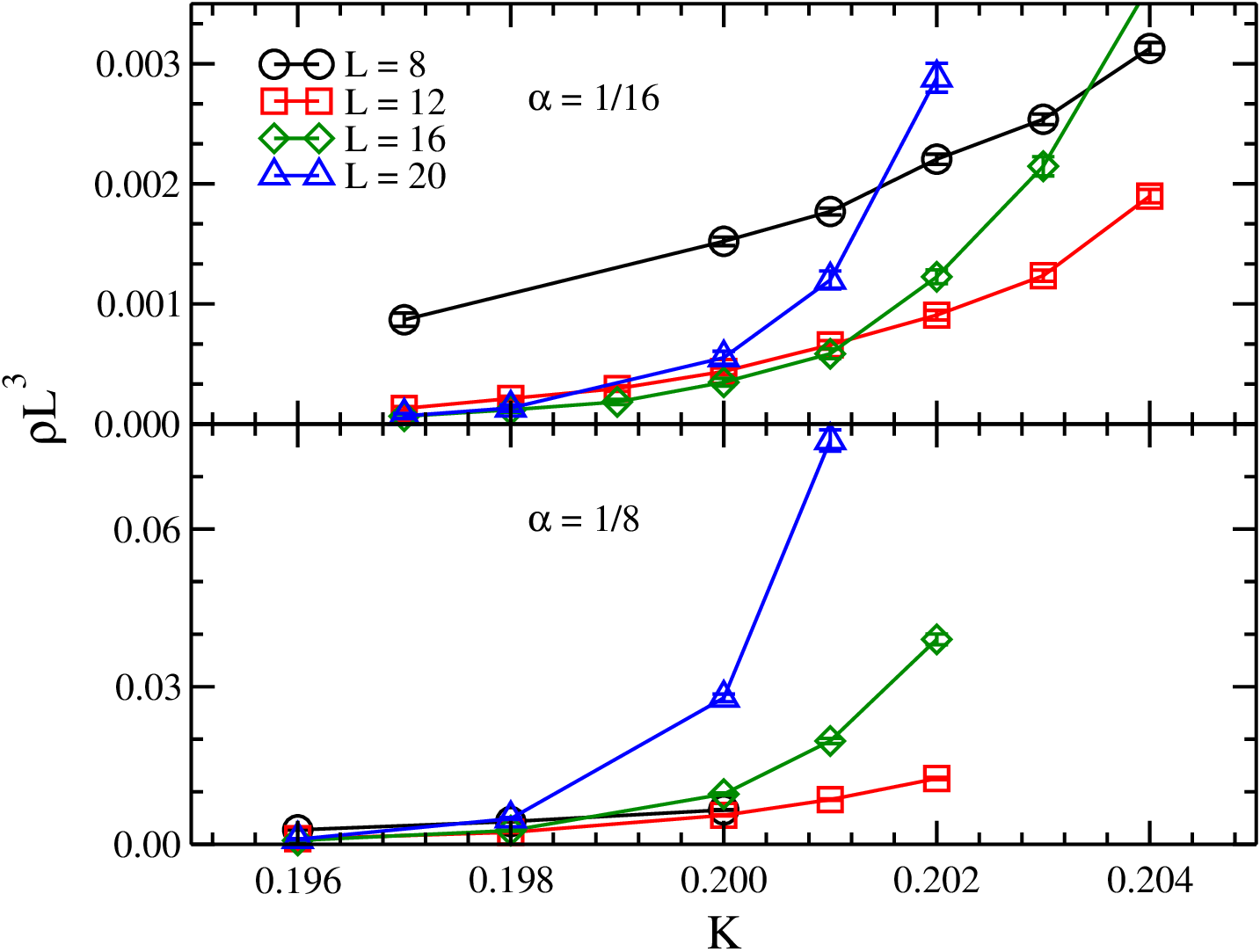}
\caption{ Search for a crossing in $\rho L^{d + z - 2}$ for $z = 2$ using lattices of size
				$L^3\times\alpha L^{z=2}$, (see also Fig.~\ref{fig:cross}). Results are shown for two different
				aspect ratios $\alpha= 1/8, 1/16$. The crossings show significant drift between different lattice
				sizes and between the two aspect ratios shown. Consequently, $z\neq 2$.} 
\label{fig:cross2}
\end{figure}

\section{\label{sec:results}Results}
We begin by a discussion of our results for the scaling of the stiffness $\rho$.  All our results
are for the three-dimensional case, hence, the scaling relation Eq.~\ref{eq:rhoscale} states that
$L^{z+1}\rho(L^{1/\nu}\delta,\alpha)$ should display a crossing at the critical point. As already
mentioned, we focus on the value of the chemical potential $\mu_{\bf r}$ distributed randomly
between $[\mu-\Delta,\mu+\Delta]$ with $\mu=\frac{1}{2}$ and $\Delta=\frac{1}{2}$.

We begin with the ansatz $z=d=3$, using lattices of size $L\times L\times L\times\alpha L^{z = 3}$.
In Fig. \ref{fig:cross} we show plots of $\rho L^4$ versus $K$ for linear system sizes ranging from
$L$ = 8 to 16. A clean crossing is observed at $K_c = 0.190(1)$ for three values of the aspect
ratio, $\alpha = 1/8$ (main panel), $\alpha = 1/32$, and $\alpha = 1/64$ (insets).  As can be seen
from Fig.~\ref{fig:cross}, our results for systems of size $L = 8$ do not scale as well as larger
system sizes. This effect is more pronounced for $L < 8$ (not shown).  This is most likely due to
corrections to scaling which for these small system sizes can not be neglected.  We also note that
an improvement in the scaling behaviour of $L = 8$ with the aspect ratio is visible in
Fig.~\ref{fig:cross}. This is presumably due to the fact that the optimal value for $\alpha$ is
given by the relation $(\xi/L)^z \sim \xi_\tau/L_\tau$~\cite{hpcs2006}.

\begin{figure}[t,floatfix]
\includegraphics[clip,width=8cm]{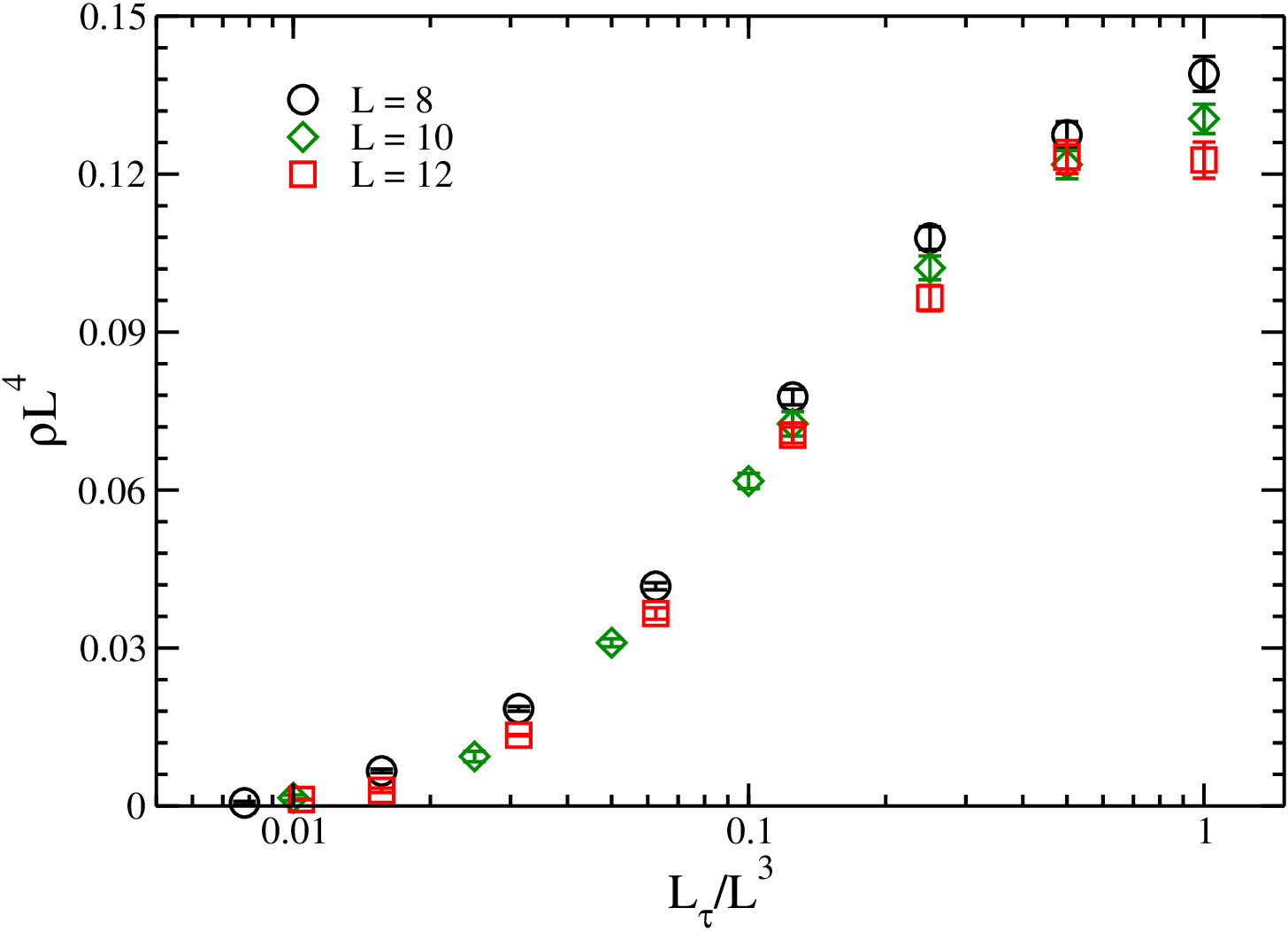}
\caption{ $\rho L^4$ versus $\alpha=L_\tau/L^z$ for system sizes $L = 8, 10, 12$ assuming $z=3$ at $K = 0.19$.
				Lattices of size $L^3\times L_\tau$ were used.  The inflection points and curvature show a very good
				collapse indicating that the dynamical critical exponent is likely $z=3$. The slight vertical drift
				with increasing $L$ likely indicates that we are slightly off the true $K_c = 0.190(1)$, within the
				bounds of our errorbars. }
\label{fig:huse} 
\end{figure}

In order to test how sensitive we are to the value of $z$ we also tried $z=2$, the value for the
dynamical critical exponent quite well established in two dimensions. Our results for this value of
$z$ are shown in Fig.~\ref{fig:cross2} for lattices of dimension $L^3\times\alpha L^{z=2}$.  For
this value of $z$ our results show significant drift in the crossing for different system sizes and
different aspect ratios as can be clearly seen in Fig.~\ref{fig:cross2}. The apparent crossing
between two system sizes are only slightly higher than the clear crossing seen using $z=3$ in
Fig.~\ref{fig:cross} but only two system sizes can be made to cross at a given $K$. From the results
in Fig.~\ref{fig:cross2} we conclude that $z\neq 2$.

As explained in section~\ref{sec:scaling} we can now hold constant the first argument of the scaling
function (\ref{eq:rhoscale}) by working at our estimate of $K_c$.  Assuming that our estimate of
$K_c=0.190(1)$ obtained with $z=3$ is correct we can now check these estimates self-consistently by
plotting $\rho L^{d+z-2=4}$ versus $L_\tau/L^{z=3}$ for a range of $L$ at $K_c=0.190(1)$.  For a
detailed explanation of the procedure we refer to Ref~\onlinecite{guo:1994:prl}.  Our results are
shown in Fig.~\ref{fig:huse}.  The curves show $\rho L^{d + z - 2=4}$ plotted for systems of linear
size $L = 8$, $10$, and $12$ as a function of the aspect ratio $\alpha=L_\tau/L^{z=3}$.  All the
curves for different $L$ and $L_\tau$ collapse onto a single curve for $z = 3$.  This result is a
rather strong confirmation that the assumption $z=3$ indeed is correct.  If one studies the curves
in detail a very slight vertical drift with increasing $L$ is noticeable. This drift is likely the
result of a small deviation, within our errorbars, of the actual critical temperature from our
estimate of $K_c=0.190(1)$.  It would have been quite interesting to study systems of linear size
$L>12$ or systems with an aspect ratio largely exceeding $L_\tau/L^z\sim 1$, however, given our
value of the dynamical critical exponent of $z=3$ this is computationally too demanding.

\begin{figure}[t,floatfix]
\includegraphics[clip,width=8cm]{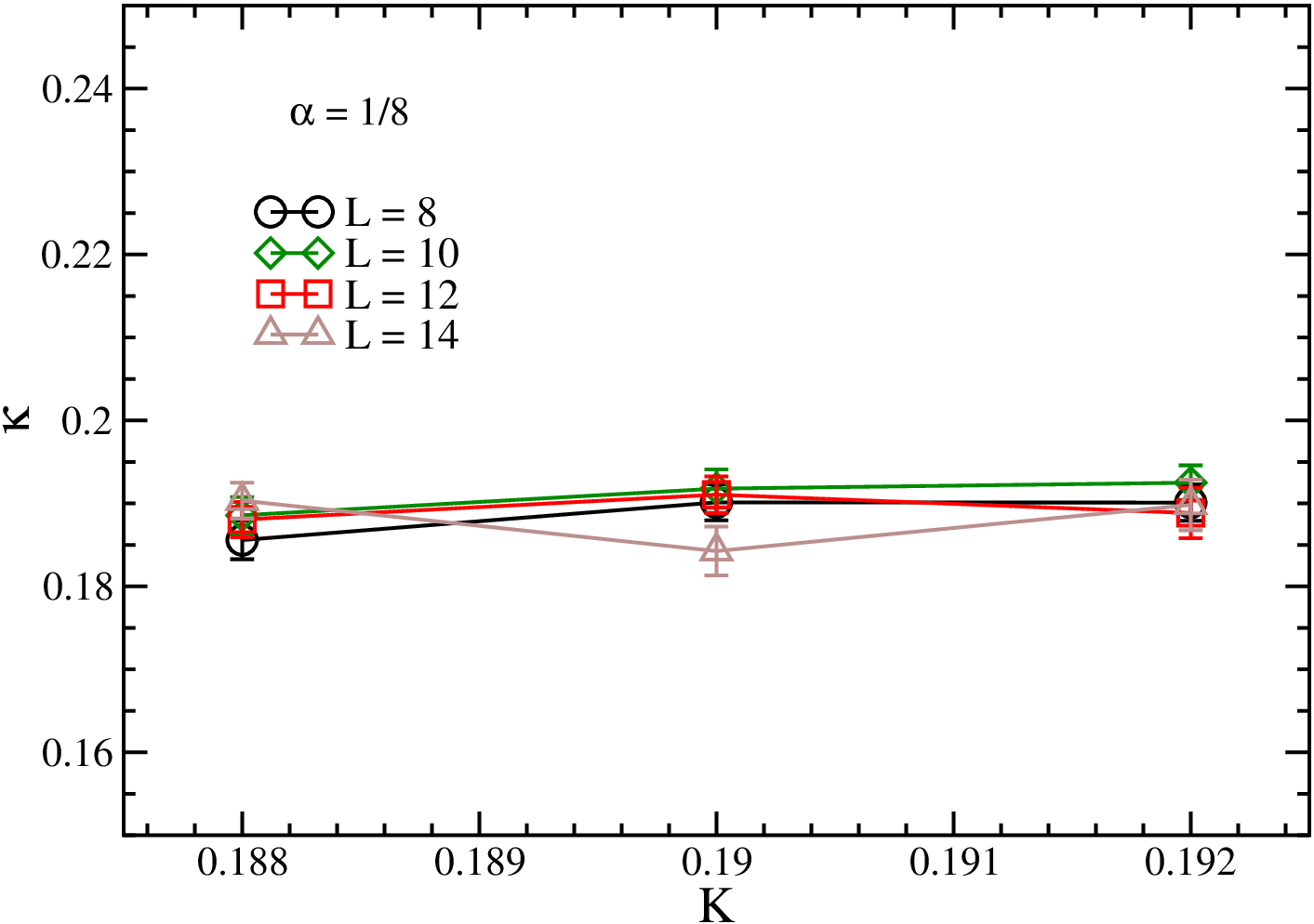}
\caption{ The compressibility, $\kappa$ shown versus $K$ near $K_c=0.190(1)$. Results are shown for $L = 8$,
				10, 12, and 14 with $\alpha = 1/8$. Note the absence of features at $K_c$ and the lack of dependence
				on system size, indicating again that $z = d = 3$.}
\label{fig:kappa}
\end{figure}

We now discuss our results for the compressibility, $\kappa$.  In Fig.~\ref{fig:kappa} the
compressibility is shown for a range of $K$ around the estimated critical point $K_c=0.190(1)$ for
several different system sizes, using the same lattice sizes as for in Fig.~ \ref{fig:cross}.  As
expected if $d = z$, the compressibility shows no dependence on the linear size of the system and is
approximately constant through the transition.  Furthermore, at $K_c$ the compressibility neither
diverges nor does it tend to zero. This is again a strong confirmation that $z=d=3$.  One should
note that the results in Fig.~\ref{fig:kappa} are only really useful for determining $z$ if the
critical point, $K_c$ is known (since one generally would expect $\kappa$ to be independent of $L$
far away from the critical point where $\xi\ll L$).  In the absence of any knowledge of $K_c$,
$\kappa$ would have to be calculated for the entire range of physically relevant $K$.  It is
therefore interesting to compare the results shown in Fig.~\ref{fig:kappa} with the attempt of
locating the critical point assuming $z=2$ shown in Fig.~\ref{fig:cross2}. The latter results show
curves intersecting pairwise around $K\sim0.2$ with the intersections moving downwards with $L$.  If
indeed the dynamical critical exponent was $z=2$ and not $z=3$ as we show here, then it would seem
very unlikely that the compressibility shown in Fig.~\ref{fig:kappa}, just below this range of
$K\sim0.2$, could be so featureless.  One would in that case have expected it to behave as
$\kappa\sim\delta^{\nu(d-z)=\nu}$, resulting in the finite-size scaling form
$\kappa=(1/L^{d-z=1})\bar\kappa(L/\xi,L_\tau/L^z)$. Since the results in Fig.~\ref{fig:kappa} are
independent of $L$ they therefore clearly exclude $z=2$.

\begin{figure}[t,floatfix]
\includegraphics[clip,width=8cm]{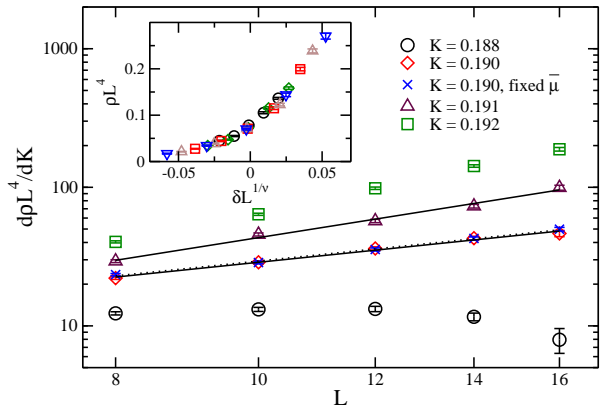}
\caption{The derivative of the stiffness, $L^4d\rho/dK$, as calculated directly during the
			simulation, plotted against the linear system size $L$. Data is shown for a range of couplings near
			$K_c$.  The solid lines indicate power-law fits to the data at $K=0.190$, with exponent $1.11$ and
			$K=0.191$ with an exponent of $1.69$. Data is also shown for simulations run in the micro-canonical
			ensemble of disorder at $K = 0.190$ (see text). The dashed line indicates a power-law fit to the
			micro-canonical data with an exponent indistinguishable from the fit to the
			canonical data. The inset shows the best collapse of the scaling curves $\rho L^4$ plotted against
			$\delta L^{1/\nu}$, with $K_c = 0.1902$ and $\nu = 0.70$. Combining these results we find $\nu =
			0.70(12)$.}
\label{fig:nu}
\end{figure}

We now turn to our results for the correlation length exponent $\nu$.  Figure~\ref{fig:nu} shows in
the main panel a plot of the derivative of the stiffness (times $L^4$) for the system sizes we
studied.  This quantity is calculated during the simulation without the use of any numerical
derivatives.  As explained in section~\ref{sec:scaling} we expect this derivative to behave as $\sim
L^{1/\nu}$ at  $K_c$.  The results for $K=0.188$ clearly deviate from a power-law consistent with
this value of $K$ being below $K_c$. From the results shown in Fig.~\ref{fig:cross} we know that
$K=0.191$ is likely above $K_c$, while $K=0.190$ is likely very close to $K_c$. Hence we fit the
results for $K=0.190$ and $K=0.191$ to a power-law, $L^4d\rho/dK = cL^{1/\nu}$, finding an exponent
of $1.11$ and $1.69$, respectively (shown as the solid lines in Fig.~\ref{fig:nu}). This allows us
to bracket the estimate of $\nu$ to the interval $0.59-0.91$. The rather broad range of this
interval is due to the fact that the relatively large value of $z$ makes this way of determining
$\nu$ extremely sensitive to a precise determination of $K_c$. We therefore combine this estimate
with a standard scaling plot of $\rho L^4$ versus $\delta L^{1/\nu}$ shown in the inset of
Fig.~\ref{fig:nu}. The best scaling plot is obtained with $\nu=0.70$ and $K_c=0.1902$ (well within
the errorbars of our estimate for $K_c$).  Combining these two estimates of $\nu$ we conclude
$\nu=0.70(12)$.  This value of $\nu$ satisfies the ``quantum" Harris criterion, $\nu \geq 2/d$,
which should hold for all disorder-driven transitions~\cite{chayes:1986:prl} and could be consistent
with this inequality being satisfied as an {\it equality}.  It also in very good agreement with the
value of $\nu = 0.69$ obtained by Herbut~\cite{herbut:2000:prb} in an expansion in powers of
$\sqrt{d-1}$ away from the lower critical dimension $d_l=1$.

\begin{figure}[t,floatfix]
\includegraphics[clip,width=8cm]{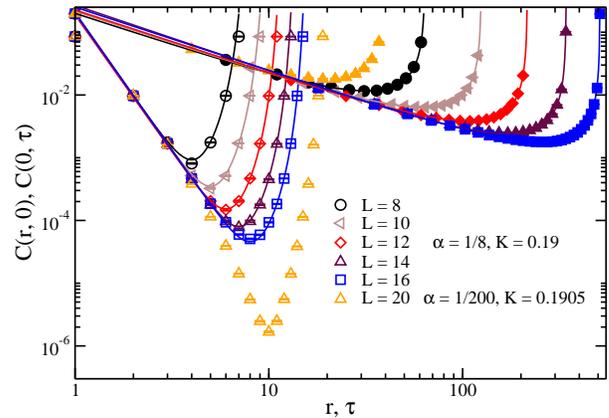}
\caption{ Spatial and temporal dependence of the particle-particle correlation function, $C(\bm{r},
				\tau = 0)$ and $C(\bm{r} = 0, \tau)$. Calculated at $K = 0.19$ and $\alpha = 1/8$ for $L = 8$ to 16.
				Solid lines are a fit to $C(\bm{r}, \tau = 0) = A(r^{-y_r} + (L - r)^{-y_r})$ and $C(\bm{r}=0, \tau
				) = A(\tau^{-y_\tau} + (L_\tau- \tau)^{-y_\tau})$. Each correlation function was fitted with the
				same power, but a unique coefficient.  Fitted values are $y_r = 4.4(4)$ and $y_\tau = 1.06(7)$.
				Also shown are curves calculated for a system of size $20^3 \times 40 (\alpha = 1/200)$ at $K =
				0.1905$. This correlation function shows a clear exponential dependence in the spatial direction,
				though the temporal dependence is consistent with the same power law decay as for systems of larger
				aspect ratio. }
\label{fig:corr}
\end{figure}

It has been suggested~\cite{pazmandi:1997:prl,pazmandi:1998:prb,bernardet:2000:prl} that the
inequality $\nu\geq 2/d$ can be violated and that in fact $\nu$ can be less than $2/d$. At the
center of this debate is the way the average over the disorder is performed. If the disorder is
chosen from a random distribution without any constraints, called the canonical ensemble of
disorder, one might ask if equivalent results are obtained if constraints are imposed on the random
distribution, the so-called micro-canonical ensemble of disorder. For instance, for the model
considered here one could constrain the random chemical potential to have {\it exactly} the same
average for each generated sample. The proof~\cite{chayes:1986:prl} of the ``quantum" Harris
criterion relies on the physically more relevant canonical ensemble of disorder being used.
Subsequent work~\cite{aharony:1998:prl,wiseman:1998:prl,igloi:1998:prb,dhar:2003:prb,%
monthus:2004:prb} have shown that even though the two ensembles in principle yield the same results
in the thermodynamic limit, their finite-size properties can in some cases be different.  All our
results have been obtained using the canonical ensemble of disorder, in that at each site a
potential was drawn from the uniform distribution $[0, 1]$. Hence, for a given sample the average
chemical potential is not exactly $1/2$.  In light of the above discussion we have therefore
performed additional simulations directly at $K_c$ but this time imposing the constraint that the
average chemical potential must be {\it exactly} $1/2$ for every disorder realization. Our results
for this micro-canonical ensemble of disorder are also shown in Fig~\ref{fig:nu} and are
indistinguishable from our results obtained with the canonical ensemble. One should note that
the uniform disorder distribution we employ is likely less sensitive to the difference between the
canonical and micro-canonical ensemble of disorder than a bimodal distribution. 
Hence we conclude that our
procedure for calculating $\nu$ is valid.

We finally discuss our results for the correlation functions which are shown in Fig.~\ref{fig:corr}
for a range of different system sizes calculated at $K_c$.  To determine the power law decay of the
correlation functions we used lattices of size $L^3\times \alpha L^z$ with $\alpha=1/8$ and $z=3$.
All the temporal correlation functions can be fitted with the same exponent $y_\tau=1.06(7)$.  We
note that this value for $y_\tau$ would appear to satisfy the equality $y_\tau\leq 1$ as an
equality.  Following section~\ref{sec:scaling} we assume that $y_\tau=(d+z-2+\eta)/z$ and using the
above determined value for $z=3$ we then infer 
\begin{equation}
\eta\sim -1.
\end{equation}
This would then satisfy Eq.~(\ref{eq:eta}) as an equality. Our results for the temporal correlation
functions have been determined out to large lattice sizes $L_\tau=512$ and appear quite stable. We
are therefore relatively confident that these results are trustworthy.

Our results for the spatial correlation functions $C({\bf r},\tau=0)$ appear less clear.  The value
we obtain for $y_r=4.4(4)$ is clearly not consistent with our other estimates for $z$, since the
ratio $y_r/y_\tau\equiv z$ then yields $z=4.15$. If one assumes that this is the correct value for
$z$ it again follows that $\eta\sim -1$.  We strongly suspect that this value of $y_r$ is due to the
relatively small spatial extent ($L\leq 16$) of the lattices used. For such small lattice sizes the
correlation functions have likely not reached their asymptotic behavior.  We have varied the
strength of the disorder, and found the same critical behaviour at $\Delta = 0.6$ and $\Delta =
1.0$.  One might attempt to reach larger lattice sizes for the spatial correlation functions by
decreasing the aspect ratio dramatically. We have attempted this by performing calculations for a
lattice of size $20^3\times 40$ at $K_c$ (also shown in Fig. \ref{fig:corr}), corresponding to an
aspect ratio of $\alpha=1/200$. In this case the spatial correlation functions show clear evidence
for an {\it exponential} decay probably due to a dimensional cross-over induced by the extremely
small aspect ratio.  In fact, it would seem to be more reasonable to {\it increase} $\alpha$ to a
more optimal aspect ratio determined by the requirement that $(\xi/L)^z\sim \xi_\tau/L_\tau$. Due to
the large value of $z$, this has proven impossible for the present study. 

\section{Conclusion}
In the present paper we have shown strong numerical evidence that the dynamical critical exponent
$z$ at the BG-SF critical point is equal to the spatial dimension for the three-dimensional
site-disordered Bose-Hubbard model. The relation $z=d$ therefore continues to hold in $d=3$.
Results for the scaling of the stiffness versus $K$, as well as at $K_c=0.190(1)$ versus $L_\tau$,
are consistent with $z=d=3$. The compressibility is almost constant and independent of system size
through the critical point consistent with this value for $z$. In addition we have obtained values
for the critical exponents $\nu=0.70(12)$ and $\eta\sim-1$, with the cautionary note that a reliable
determination of $\eta$ is impeded by the very small spatial extent of the lattices available.
In light of the recent work by Bernardet {\it et al.}~\cite{bernardet:2000:prl}, it would have been
very interesting to analyze each disorder realization seperately, 
in order to determine a characteristic $K_i$ for each sample. The scaling analysis should then
be redone using $K_i$. Unfortunately, we were not able to perform such an analysis for this work.

\acknowledgments
All our calculations have been performed at the SHARCnet computational facility and we thank NSERC
and CFI for financial support.

\bibliography{bhd3d}

\end{document}